\documentclass[namedreferences,hyperref,optionalrh]{spr-sola}

\usepackage{geometry}
\usepackage{rotating}
\usepackage{graphicx}

\usepackage{multirow}
\usepackage{appendix}
\usepackage{threeparttable}
\usepackage{bm}
\usepackage{ragged2e}

\newcommand{\aap}{{\it Astron. Astrophys.}}

\newcommand{\apj}{{\it Astrophys. J.}}
\newcommand{\apjl}{{\it Astrophys. J. Lett.}}

\newcommand{\grl}{{\it Geophys. Res. Lett.}}

\newcommand{\jgr}{{\it J. Geophys. Res.}}

\newcommand{\nat}{{\it Nature}}

\newcommand{\solphys}{{\it Sol. Phys.}}
 
\newcommand{\ssr}{{\it Space Sci. Rev.}} 
\chardef\us=`\_

\begin{document}
\begin{frontmatter}

\title{Influence of the Deformation of Coronal Mass Ejections on Their In-Situ Fitting with Circular-Cross-Section Flux Rope Models}

\author[addressref=aff1, corref, email={bin.zhuang@unh.edu}]{\inits{B.}\fnm{Bin}~\snm{Zhuang}\orcid{0000-0002-5996-0693}}
\author[addressref=aff1, email={noe.lugaz@unh.edu}]{\inits{N.}\fnm{No\'{e}}~\snm{Lugaz}\orcid{0000-0002-1890-6156}}
\author[addressref=aff1, email={nada.alhaddad@unh.edu}]{\inits{N.}\fnm{Nada}~\snm{Al-Haddad}\orcid{0000-0002-0973-2027}}
\author[addressref=aff1, email={charlie.farrugia@unh.edu}]{\inits{C.}\fnm{Charles J.}~\snm{Farrugia}\orcid{0000-0001-8780-0673}}
\author[addressref=aff2, email={ute.amerstorfer@geosphere.at}]{\inits{U.}\fnm{Ute}~\snm{Amerstorfer}\orcid{0000-0003-1516-5441}}
\author[addressref=aff2, email={emma.davies@geosphere.at}]{\inits{E.}\fnm{Emma~E.}~\snm{Davies}\orcid{0000-0002-1890-6156}}
\author[addressref=aff3, email={manuela.temmer@uni-graz.at}]{\inits{M.}\fnm{Manuela}~\snm{Temmer}\orcid{0000-0003-4867-7558}}
\author[addressref=aff2, email={hannah@ruedisser.at}]{\inits{H.}\fnm{Hannah T.}~\snm{R\"udisser}\orcid{0000-0002-2559-2669}}
\author[addressref=aff1, email={wenyuan.yu@unh.edu}]{\inits{W.}\fnm{Wenyuan}~\snm{Yu}\orcid{0000-0002-2917-5993}}
\author[addressref=aff4, email={tingyu.gou@cfa.harvard.edu}]{\inits{T.}\fnm{Tingyu}~\snm{Gou}\orcid{0000-0003-0510-3175}}
\author[addressref=aff1, email={reka.winslow@unh.edu}]{\inits{R.}\fnm{R\'{e}ka M.}~\snm{Winslow}\orcid{0000-0002-9276-9487}}

\address[id=aff1]{Institute for the Study of Earth, Oceans, and Space, University of New Hampshire, Durham, NH, USA}
\address[id=aff2]{Austrian Space Weather Office, GeoSphere Austria, 8020 Graz, Austria}
\address[id=aff3]{Institute of Physics, University of Graz, A-8010 Graz, Austria}
\address[id=aff4]{Center for Astrophysics $|$ Harvard \& Smithsonian, Cambridge, MA 02138, USA}

\justifying


\begin{abstract}
Understanding the properties, especially the magnetohydrodynamic (MHD) invariants, of coronal mass ejections (CMEs) measured in-situ is key to bridging the CME properties from the Sun to interplanetary space. In order to investigate CMEs from the in-situ measurements that provide a one-dimensional (1-D) cut of the CME parameters over the spacecraft trajectory, various magnetic flux rope (MFR) models have been developed, among which the models with a circular cross-section are the most popular and widely used. CMEs are found to be deformed during their propagation in interplanetary space, in which the cross-section may be flattened in the direction of propagation, i.e., to develop an elliptical or even pancake-like shape. We use numerical MHD simulations in 2.5-D to investigate the influence of the CME deformation on the in-situ fitting using two linear force-free MFR models with a circular cross-section, and we focus on the axial and poloidal magnetic fluxes, which are conserved in the ideal MHD frame and simulations. We quantitatively compare the fitted axial and poloidal fluxes with those in simulations. We find that both models underestimate the axial flux compared to that in simulations, and such underestimation depends on the CME deformation. However, the fitting of the poloidal flux is independent of the deformation. We discuss the reasons for the axial flux underestimation and the implication of the CME deformation for the CME in-situ fitting. 
\end{abstract}
\end{frontmatter}

%

\section{Introduction} \label{sec: intro}
Coronal mass ejections (CMEs) are large-scale solar transients that erupt from the corona, propagate to interplanetary space, and provide a large amount of magnetic flux into the heliosphere. When a CME passes a spacecraft in interplanetary space, in-situ measurements taken by the spacecraft provide a time series of measurements, which is often equated to a one-dimensional (1-D) cut of the CME parameters along the spacecraft trajectory under the assumption of CMEs as static structures. This assumption has been recently analyzed in more detail, both using data \citep{regnault2023,regnault2024a} and simulations \citep{regnault2024b}. For CMEs measured in-situ, there exists a typical subset called magnetic clouds (MCs) that have observational properties of a) an enhanced magnetic field strength, b) a smooth rotation of the magnetic field vector through a large angle, and c) a low proton temperature \citep{burlaga1981}. MCs, serving as one of the most important aspects of CME studies, have been investigated widely and in-depth since the 1980s.

To obtain the 2-D or 3-D configuration of MCs from the 1-D in-situ data, a variety of techniques have been continuously developed and improved. Based on the concept that MCs consist of twisted magnetic flux rope (MFR) structures \citep{burlaga1981}, many MFR models, either force-free \citep{goldstein1983,marubashi1986,burlaga1988} or non-force-free \citep{hidalgo2002}, have been proposed to describe the MC magnetic field configurations. In those models, the assumption of the geometry of the MFR is a crucial point. Those geometries include circular cylindrical symmetry, non-circular cylindrical symmetry, and/or toroidal symmetry. In addition, the Grad-Shafranov reconstruction was developed to incorporate the magnetostatic equilibrium with an invariant direction of MCs and to recover the MC magnetic field configuration, without any presuppositions of the MC magnetic field descriptions or cross-section \citep{hu2001}. One can refer to the review papers of, e.g., \cite{forbes2006}, \cite{al-haddad2013}, and \cite{zhang2021} for more details about different techniques. Even though there have been many advanced techniques, the techniques that are popular and widely used in the heliophysics and space weather communities are still the models with a circular cross-section that have concise solutions and can be easily applied to in-situ measurements \citep[e.g.,][]{wang2015,wang2016,wang2018,nieves2016,yu2022}.

The application of these techniques provides the CME parameters when the CME is measured locally and further helps bridge the CME properties from the Sun to interplanetary space \cite[e.g.,][]{qiu2007,mostl2009,hu2014,wang2017,pal2017,wang2018}, especially those invariants, e.g., magnetic flux in the ideal magnetohydrodynamics (MHD) frame or magnetic helicity. For example, \cite{qiu2007} investigated the magnetic flux budget of CMEs in association with magnetic reconnection in the low corona and compared the flux components (poloidal and axial) measured in-situ with the reconnection flux estimated in the corona. They found that the CME poloidal flux estimated in-situ is comparable with the reconnection flux. Similarly, \cite{hu2014} found that the amount of the twisted flux per unit length of CMEs is comparable with the total reconnection flux on the Sun. \cite{wang2017} studied the twist distribution inside the CME-associated MFR estimated both in the corona and in-situ, in which the MFR twist in the corona was estimated based on the ratio of the poloidal magnetic flux to the axial magnetic flux. 

In order to have reliable comparisons between solar and in-situ estimates, in addition to addressing the uncertainties in the reconnected flux from the flare respective \citep{temmer2017}, a key requirement is to accurately fit or reconstruct CME properties. Uncertainties of the fitting techniques need to be addressed. \cite{al-haddad2013} found that different techniques may have different outputs even though they are applied to the same CME. Furthermore, the evolution, e.g., magnetic erosion \citep{dasso2006,ruffenach2012,lavraud2014,wang2018,farrugia2023} or deformation of the CME during its propagation are expected to affect the fitted results. In this study, we focus on the deformation aspect. While propagating in interplanetary space, the cross-section of CMEs is found to be deformed (or flattened) in the propagation direction from an initial circular shape to an elliptical shape or even a pancake-like shape \cite[e.g.,][]{manchester2004,riley2004a,owens2006,savani2011}. Therefore, employing the popular circular cross-section model for a deformed CME may lead to inaccurate estimations of CME parameters, and comparing those MHD invariants of CMEs between solar and in-situ measurements is then affected. To investigate the influence of the deformation, it is most effective to use numerical simulations.

The development of numerical methods enables us to study the eruption and propagation of CMEs from the Sun to the heliosphere \cite[see, e.g.,][]{odstrcil1999,manchester2004,manchester2017,lugaz2009,feng2020,shen2022}. Past studies have shown that numerical simulations can be used to test the output of the fitting techniques \cite[e.g.,][]{riley2004,vandas2010,al-haddad2011,al-haddad2019,lynch2022}. For example, \cite{riley2004} used blind tests to compare different fitting techniques of the simulated CME passing through in-situ spacecraft along different trajectories. Furthermore, the reconstruction and fitting techniques were applied to predominantly writhed CMEs and twisted MFRs to test their capability to reconstruct either the writhed or twisted magnetic field configuration \citep{al-haddad2011,al-haddad2019}. It was found that due to the appearance of magnetic field rotation resulting from writhed magnetic field lines, the twisted MFR models can inadvertently reconstruct a non-twisted structure.  

In this paper, we use numerical simulations to investigate the influence of the flattened cross-section of the (CME-associated) MFR on the fitted results using models with a circular cross-section. We focus on magnetic flux, which is conserved during the CME propagation in our numerical scheme. We introduce the numerical method and fitting techniques in Section~\ref{sec: simu_fit}. In Section~\ref{sec: result}, we present comparisons between the MFR parameters obtained from simulations and fitting techniques. Discussion and conclusions are provided in Sections~\ref{sec: dis} and \ref{sec: con}, respectively.

\section{Numerical Setup and MFR Fitting Model} \label{sec: simu_fit}
\subsection{Numerical Simulation}
Following \cite{sun2005} and \cite{zhuang2018}, we solve the MHD equations in spherical coordinates ($r, \theta, \phi$) and consider 2.5-D problems in the heliospheric meridional plane ($\frac{\partial}{\partial \phi}=0$). 2.5-D simulations are suitable for the motivation of this study for the following three reasons. First, the used fitting techniques also solve 2.5-D problems, i.e., they assume symmetry along the MFR axis. Second, the MFR boundaries can be accurately identified in 2.5-D simulations, while the input of the MFR boundaries was also found to influence the fitted results \citep{riley2004}. Third, it is relatively easier to achieve the conservation of magnetic fluxes and helicity in a 2.5-D simulation as compared to a 3-D simulation.

The ideal 2.5-D MHD equations are listed as follows.
\begin{equation}\label{eq1}
\frac{\partial \rho}{\partial t}+\nabla \cdot \left (\rho \bm{v} \right ) =0,
\end{equation}
\begin{equation} \label{eq2}
\frac{\partial \bm{v}}{\partial t} + \bm{v} \cdot \nabla \bm{v} + R \nabla T + \frac{RT}{\rho} \nabla \rho + 
\frac{1}{\mu_0 \rho} [\mathcal{L} \psi \nabla \psi + \bm{B}_{\phi} \times (\nabla \times \bm{B}_{\phi})] + \frac{1}{\mu_0 \rho r \sin \theta} \nabla \psi \cdot (\nabla \times \bm{B}_{\phi}) \hat{\phi} + \frac{GM_s}{r^2} \bm{\hat{r}} = 0,
\end{equation}
\begin{equation} \label{eq3}
\frac{\partial \psi}{\partial t}+\bm{v} \cdot \nabla \psi=0,
\end{equation}
\begin{equation} \label{eq4}
\frac{\partial \bm{B_\phi}}{\partial t}+r \sin \theta \nabla \cdot \left (\frac{\bm{B_\phi} \bm{v}}{r \sin \theta} \right)+ \left [\nabla \psi \times \nabla \left (\frac{\bm{v_\phi}}{r \sin \theta} \right) \right]_ \phi =0,
\end{equation}
\begin{equation}\label{eq5}
\frac{\partial T}{\partial t}+\bm{v} \cdot \nabla T+(\gamma -1)T \nabla \cdot \bm{v}=0.
\end{equation}
$\rho$, $\bm{v}$, $T$, $\psi$, and $\bm{B}$ are density, flow velocity, temperature, magnetic flux function, and magnetic field, respectively. $\bm{B}$ is expressed as follows:
\begin{equation} \label{eq6}
\bm{B}= \nabla \times \left (\frac{{\psi}}{r \sin \theta} \hat{\phi}\right)+\bm{B_ \phi}, \qquad \bm{B_ \phi}=B_ \phi \hat{\phi},
\end{equation}
and the operator $\mathcal{L}$ acts as:
\begin{equation} \label{eq7}
\mathcal{L} \equiv \frac{1}{r^2 \sin^2 \theta} (\frac{\partial^2 }{\partial r^2}+\frac{1}{r^2} \frac{\partial^2 }{\partial \theta^2}-\frac{\cot \theta}{r^2} \frac{\partial }{\partial \theta}).
\end{equation}
$R$ is the gas constant, $\mu_0$ is the vacuum magnetic permeability, G is the gravitational constant, $M_s$ is the mass of the Sun, and $\gamma$ is the polytropic index.

The numerical units in the simulations are introduced in \cite{zhuang2018} in detail. Different from \cite{sun2005} and \cite{zhuang2018}, in this study, we extend the simulation domain in the radial direction to 645~$R_\odot$ to reduce the numerical effect caused by the outer boundary, while we require the CME rear to propagate to at least 260~$R_\odot$, adopt a distance-dependent polytropic index, and adjust the magnetic field strength at the coronal base to be 1.5~G. The computational domain is thus taken as $1 \le r \le 645$~$R_\odot$ ($r=1$~$R_\odot$ represents the coronal base) and $0^\circ \le \theta \le 90^\circ$, discretized into $264\times 92$ grid points and symmetric with respect to the equatorial plane ($\theta=90^\circ$). The radial grid spacing is set to be a) increasing according to a geometric series of a common ratio of 1.02 between 1 and 10~$R_\odot$, b) uniform (0.625) between 10 and 30~$R_\odot$, and c) increasing according to a ratio of 1.02 between 30 and 645~$R_\odot$. In past studies, the polytropic index $\gamma$ is set as a constant of 1.05, aiming to add extraneous heating of the solar corona and provide supersonic solar wind solutions. However, when the computational domain is extended, such a value introduces significantly more heating at larger distances, and is inconsistent with measurements. Based on the estimates of $\gamma$ in some past studies \cite[e.g.,][]{totten1995,nicolaiu2023}, in our simulations, $\gamma$ is set to be 1.05 between 1 and 30~$R_\odot$, to linearly increase to 1.42 between 30 and 215~$R_\odot$, and to be 1.42 after 215~$R_\odot$. As such, we find that the solar wind and CME parameters obtained in-situ are comparable with real measurements; obtaining a more realistic variation of $\gamma$ \cite[e.g.,][]{roussev2003} is outside the scope of this study.

The simulation is performed in the following sequences (see Figure~\ref{fig: mfrcontour}). First, a steady solar wind background with a helmet streamer astride the equator is constructed. Second, a MFR emerges from the streamer base by adjusting the inner boundary conditions and stays beneath the streamer in equilibrium with three initial parameters of total poloidal magnetic flux ($\Phi_p$), total axial magnetic flux ($\Phi_z$), and total mass ($M$). Third, the eruption of the MFR is triggered by adjusting the three parameters to cause the so-called catastrophe of the coronal system \cite[e.g.,][]{forbes1991,sun2005,zhuang2018,zhuang2019,zhang2020}. During the propagation of the MFR, these three parameters are conserved. In this study, we allow the occurrence of numerical magnetic reconnection to avoid numerical cases with a long-stretching current sheet beneath the MFR at larger distances. Due to the usage of the magnetic flux function, the identification of the initial \cite[indicating the stage just at the catastrophic point;][]{zhuang2018} MFR boundary can be performed accurately and is not affected by the introduction of numerical reconnection occurring inside the current sheet region beneath the MFR. In addition, this numerical reconnection does not introduce additional axial magnetic flux. Therefore, the reconnected fields are not regarded as part of the MFR, and the conservation of parameters of the initial MFR remains unaffected. In the following part, we focus on the parameters of the initial MFR. 

We obtain the synthetic in-situ profiles of the magnetic field and proton parameters at different locations, i.e., at $r=15$, 40, 80, 120, 160, 215 (equivalent to 1~au), and 260~$R_\odot$, and with an angular separation of $0^\circ$, $4^\circ$, and $9^\circ$ from the equator (see the three dashed straight lines in Figure~\ref{fig: mfrcontour}). We have 21 synthetic spacecraft in total. The parameters are in a coordinate system with the origin at the spacecraft and the three directions aligned along $r$, $\theta$, and $\phi$. The three axes form the local cartesian coordinates. It is intuitive at $\Delta \theta=0^\circ$, and for simplicity, we temporarily call it ``Sun-Earth'' line. Under this assumption, $r$ is related to the ``Sun-Earth'' line, $\theta$ is perpendicular to the ecliptic plane, and $\phi$ completes the right-hand coordinate system. The positions of the synthetic spacecraft at $\Delta \theta=4^\circ$ or $9^\circ$ are equivalent to the scenario in which the propagation direction of the MFR is $4^\circ$ or $9^\circ$ relative to the ``Sun-Earth'' line in latitude while the spacecraft remains on the ``Sun-Earth'' line, or the helmet streamer is tilted in latitude as compared to the ecliptic plane.

\subsection{MFR Fitting Model}
We use two MFR fitting techniques with a circular cross-section here. The first is the velocity-modified cylindrical linear force-free flux rope model \citep{wang2015}. This model uses the Lundquist solution \citep{lundquist:1950} to describe the MFR magnetic field components and incorporates the measured velocity components in the fitting by considering the propagation and expansion of the MFR as well as the plasma poloidal motion inside the MFR.  The three magnetic field components in the so-called MFR coordinates ($r, \varphi, z$) (here, the three axis vectors are different from those in simulations) are described as follows:
\begin{equation} \label{lundquist_eq1}
    \bm{B_r} =0,
\end{equation}
\begin{equation} \label{lundquist_eq2}
    \bm{B_\varphi} = HB_0J_1(\alpha r) \bm{\hat{\varphi}},
\end{equation}
\begin{equation} \label{lundquist_eq3}
    \bm{B_z} = B_0J_0(\alpha r) \bm{\hat{z}},
\end{equation}
in which $H=\pm 1$ is the handedness (sign) of the MFR helicity, $B_0$ is the magnetic field strength at the MFR axis, $\alpha=2.41/R_0$ is the constant force-free factor, $R_0$ is the radius of the MFR cross-section, and $J_0$ and $J_1$ are the zero-order and first-order of the Bessel functions. When incorporating the MFR radial expansion into the model, $B_0$ becomes time-dependent and expressed as follows:
\begin{equation}\label{eq: lundquist_b0t}
    B_0(t)=B_0(t_0) \left [ \frac{R(t_0)}{R(t)} \right] ^2,    
\end{equation}
based on the magnetic flux conservation, following \citet{farrugia1992}. In Equation~\ref{eq: lundquist_b0t}, $t_0$ is the time when the synthetic spacecraft first encounters the MFR. We note that in practice, unlike the model of \citet{yu2022} described below, $t_0$ is not used during the fitting as $t-t_0$ can also represent the time relative to the MFR front boundary. $R(t)=R(t_0)+v_e(t-t_0)$, and $v_e$ is a constant expansion speed at the MFR boundary. $v_e$ can also be fitted with two additional parameters $\bm{v_c}=(v_r, \ v_\phi, \ v_\theta)$  by incorporating the velocity profiles and separating the MFR bulk velocity ($v_r$) and the poloidal speed at the MFR boundary ($v_p$). See \citet{wang2015} for details.

We note that the MFR has a global toroidal symmetry based on the numerical solution in spherical coordinates, which differs from the fitting model that assumes a cylindrical symmetry (straight axis). However, when fitting local measurements, we consider that the cylindrical model is valid since the torus is locally approximated by a straight tube if the MFR radius is much smaller than the heliocentric distance. In the simulations, at the six selected distances, the ratio of the MFR radius to the distance varies from 0.20 at 20~$R_\odot$ to 0.08 at 260~$R_\odot$, which supports the use of cylindrical models. Furthermore, the model with the Lundquist solution and expansion considered is suitable for the simulated case because a) the axial magnetic field component at the MFR boundaries is zero (Equation~\ref{lundquist_eq2}), and b) the MFR has expansion along the radial direction as shown by the synthetic in-situ velocity profiles. 

In brief, this model fits all parameters simultaneously and determines the final outputs based on the minimum normalized root-mean-square ($\chi$). The fitting progress of the model and the output parameters are described in \cite{wang2015}. In this study, we focus on the parameters $B_0$, $R_0$, $\theta_0$, $\phi_0$, $\Phi_p$, and $\Phi_z$. $B_0$ and $R_0$ are determined when the MFR front first reaches the spacecraft. $\theta_0$ and $\phi_0$ indicate the direction of the MFR axis in the local simulation coordinates. $\Phi_p$ and $\Phi_z$ are the poloidal and axial magnetic fluxes of the MFR, which are expressed as follows:
\begin{equation} \label{lundquist_eq4}
    \Phi_p = \int_{0}^{l} \int_{0}^{R_0} B_\varphi dr dz = 0.416B_0 R_0 l,
\end{equation}
\begin{equation} \label{lundquist_eq5}
    \Phi_z = \int_{0}^{2\pi} \int_{0}^{R_0} B_z r dr d\varphi = 1.35B_0 R_0^2,
\end{equation}
where $l$ is the length of the MFR axis. Following \cite{wang2015}, we set $l$ as $l=\frac{\pi+2}{2}D_a$, in which $D_a$ is the heliocentric distance of the MFR axis when the MFR front reaches the spacecraft. In order to have a consistent comparison of $\Phi_p$, the simulated $\Phi_p$ needs to be multiplied by a factor of 2.44 ($2\pi / \frac{\pi+2}{2}$). The simulated and fitted MFR parameters are also listed in Table~\ref{tb: table}, including the fitted impact parameter ($d$; the sign of $d$ is not considered due to the symmetry of the MFR with respect to the equator).

The second model is the radially-expanding Lundquist model \citep{yu2022} inherited from the development of the expanding Lundquist solution \citep{farrugia1992,farrugia1993}, with the magnetic field components in the MFR coordinates expressed as follows:
\begin{equation} \label{exlundquist_eq1}
    \bm{B_r} =0,
\end{equation}
\begin{equation} \label{exlundquist_eq2}
    \bm{B_\varphi}(t) = H(B_0/\tau)J_1(\alpha r/\tau) \bm{\hat{\varphi}},
\end{equation}
\begin{equation} \label{exlundquist_eq3}
    \bm{B_z}(t) = (B_0/\tau^2)J_0(\alpha r/\tau) \bm{\hat{z}},
\end{equation}
in which $\alpha=2.41/R_0$, $\tau = 1+t/t_0$, and $t_0$ represents the MFR self-similar expansion time during its transit from the Sun to the spacecraft. Parameter $t_0$ is estimated by fitting the decreasing velocity profile with Equation~\ref{farrugia_eq}:
\begin{equation} \label{farrugia_eq}
    V_r(t) = (U+R'_0/t_0)/(1+t/t_0),
\end{equation}
where $U$ is the velocity at the location with the maximum magnetic field strength (i.e., the MFR axis), and $R'_0$ indicates the radius of the MFR when first encountered by the synthetic spacecraft. Different from the first model, the second one inputs initial estimates of $\theta_0$ and $\phi_0$ using minimum variance analysis into the fitting and further fits $B_0$, $\theta_0$, $\phi_0$, and $d$. $t_0$ is fixed during the fitting, and $R'_0$ is replaced by $R_0$ which is derived based on the average solar wind speed within the MFR period and the fitted $d$. We note that at $t=0$ of the MFR front boundary, Equations~\ref{exlundquist_eq2} and \ref{exlundquist_eq3} simplify to Equations~\ref{lundquist_eq2} and \ref{lundquist_eq3}, and thus Equations~\ref{lundquist_eq4} and \ref{lundquist_eq5} are also applicable to the calculations $\Phi_p$ and $\Phi_z$ in the expanding Lundquist model.

In this study, we do not aim to compare the fitted qualities between these two models. Our focus is to compare $\Phi_p$ and $\Phi_z$ obtained between simulations and fittings.

\section{Result}\label{sec: result}
\subsection{Simulation and Fitting}
\begin{figure}[!hbt]
    \centering
    \includegraphics[width=0.8\textwidth]{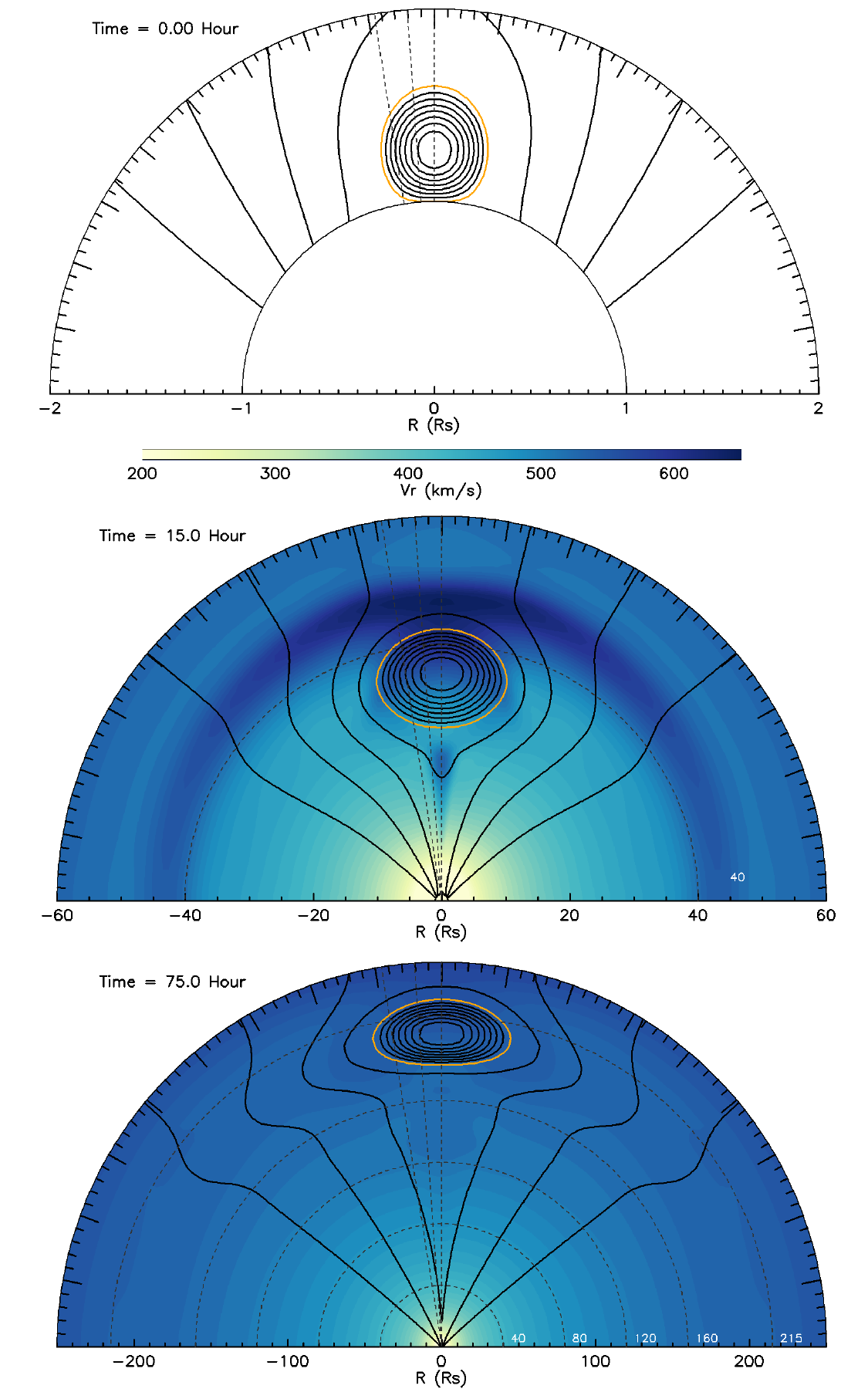}
    \caption{MFR before (top panel) and after (middle and bottom panels; at two different time steps) the eruption in the meridional plane. The solid curves depict magnetic field lines in the simulation domain. The orange circle marks the MFR boundary. In the middle and bottom panels, the contours are filled with the radial velocity information. The three dashed straight lines indicate the latitudes of the synthetic spacecraft.
    }
    \label{fig: mfrcontour}
\end{figure}

Figure~\ref{fig: mfrcontour} shows the MFR before (top panel) and after (middle and bottom panels) the eruption. To trigger the eruption, the MFR parameters are adjusted to be $\Phi_p=4.04\times 10^{21}$~Mx per radian, $\Phi_z=2.42\times 10^{21}$~Mx, and $M=1.07\times 10^{13}$~kg per radian. We have tested different sets of the MFR initial parameters, and the case presented in this paper is used because a) the synthetic in-situ profiles show that the ratio of the maximum axial magnetic field vector to the maximum poloidal field vector is close to the ratio described by the Lundquist solution (Equations~\ref{lundquist_eq2} and \ref{lundquist_eq3}) and b) the MFR has an approximately linear-decreasing in-situ velocity profile indicating its radial expansion. The solid curves and circles in the figure depict the magnetic field lines of the background and MFR in the meridional plane, respectively.
The orange circle outlines the accurate MFR boundary determined by the magnetic flux function. In the middle and bottom panels, the contours are filled with radial velocity values. It is found that the MFR becomes flattened to an elliptical cross-section during its outward propagation and drives a high-velocity compression region ahead of itself (middle panel).

\begin{figure}[!hbt]
    \centerline{\includegraphics[width=1.\textwidth,clip=]{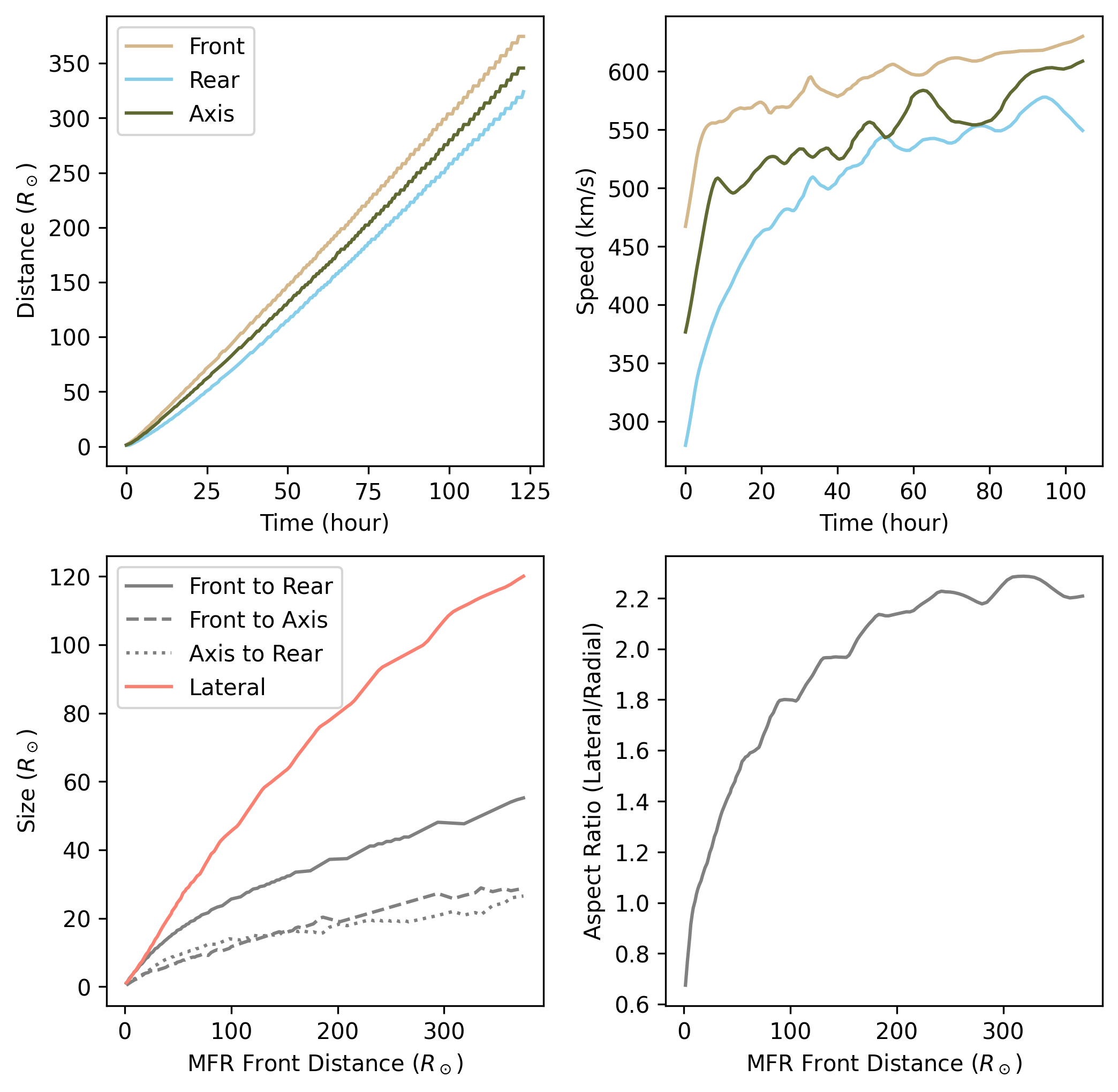}}
    \caption{Variations of the MFR propagation and geometric parameters. The top left and right panels display the temporal variations of the heliocentric distance and radial velocity of the front, axis, and rear of the MFR, respectively. The bottom left panel shows the distance variations of the MFR radial sizes from the front to rear (solid), front to axis (dashed), and axis to rear (dotted), as well as the lateral size (red). The bottom right panel displays the variation of the aspect ratio of the MFR cross-section.
    }
    \label{fig: aspect_ratio}
\end{figure}
Figure~\ref{fig: aspect_ratio} shows the variations of the MFR propagation and geometric parameters. The top left and right panels display the temporal variations of the heliocentric distance and radial velocity of the front, axis (determined by location with the maximum magnetic flux function), and rear of the MFR, respectively. It is found that the MFR experiences an impulsive acceleration within the first 6 hours and a gradual but weaker acceleration contributed by numerical reconnection at larger distances. After $\sim$60 hours, the velocities of the front, axis, and rear of the MFR become nearly constant. The velocity differences of the MFR substructures indicate the existence of expansion in the radial direction. The bottom left panel shows the variations of the MFR radial sizes from the front to the rear (solid), from the front to the axis (dashed), and from the axis to the rear (dotted), as well as the lateral size (red) along with the distance of the MFR front. The radial sizes from the front to the axis and from the axis to the rear are similar during the MFR propagation, which indicates that the axis is roughly located at the geometric center of the MFR in the radial direction. We estimate the rate of increase in the radial size using a power-law distribution following past studies \cite[e.g.,][]{bothmer1998,savani2009,gulisano2010,nieves2013,zhuang2023}. The radial size between the front and the rear shows a power-law index of 0.82 within $\sim$20~$R_\odot$ and 0.62 beyond. The result of 0.82 is similar to the power-law indices found in studies such as \cite{bothmer1998} and \cite{gulisano2010}, which are around 0.8, while the result of 0.62 is similar to the estimates in \cite{savani2009} (0.55--0.65). Similarly, using 3-D simulations, \cite{al-haddad2019} found that the power-law index for the CME radial size increase is around 0.7 from the corona to 0.5~au. The decrease in the expansion rate is due to the fact that the rear part of the MFR continues experiencing a stronger acceleration compared to the front part, as shown in the top right panel of Figure~\ref{fig: aspect_ratio}. Focusing on the radial size from the front to the axis, the power-law index is found to be around 0.8 during the propagation of the MFR.

The lateral size is the size along a line perpendicular to the Sun-MFR axis line and intersecting the leftmost and rightmost edges of the MFR. Finally, the distance variation of the aspect ratio (lateral size to the radial size from front to rear) of the MFR is given in the bottom right panel of Figure~\ref{fig: aspect_ratio}. The aspect ratio is found to increase during the MFR cross-section propagation and reach 1.0 and $\sim$2.1 when the MFR front is at 15 and 215~$R_\odot$, respectively. The aspect ratio increases much more slowly after around 150~$R_\odot$ and remains approximately constant beyond 215~$R_\odot$. Even though the aspect ratio is increasing, we find that the angular width decreases during the MFR propagation, from $34^\circ$ initially to $24^\circ$ at 215~$R_\odot$. Decreases in the CME angular width during the CME propagation have been found in past studies such as \cite{lugaz2010} and \cite{nieves2013}. Magnetic tension force may play a role in decreasing the angular width (rather than maintaining a constant angular width) and keeping the MFR cross-section nearly circular, thus preventing the aspect ratio from being too large \citep{suess1988}. 

The distance variation of the aspect ratio enables us to estimate the influence of the flattened MFR cross-section on the fitting if the synthetic spacecraft are set at different distances. We do so for the synthetic in-situ parameter profiles at 21 different locations. 
We also test the influence of the impact parameter (the closest approach of the observational path to the MFR axis) on the fitting, and thus three different Sun-spacecraft lines are used with angular separations ($\Delta \theta$) of $0^\circ$, $4^\circ$, and $9^\circ$.
\begin{figure}[!hbt]
    \centering
    \includegraphics[width=0.82\textwidth]{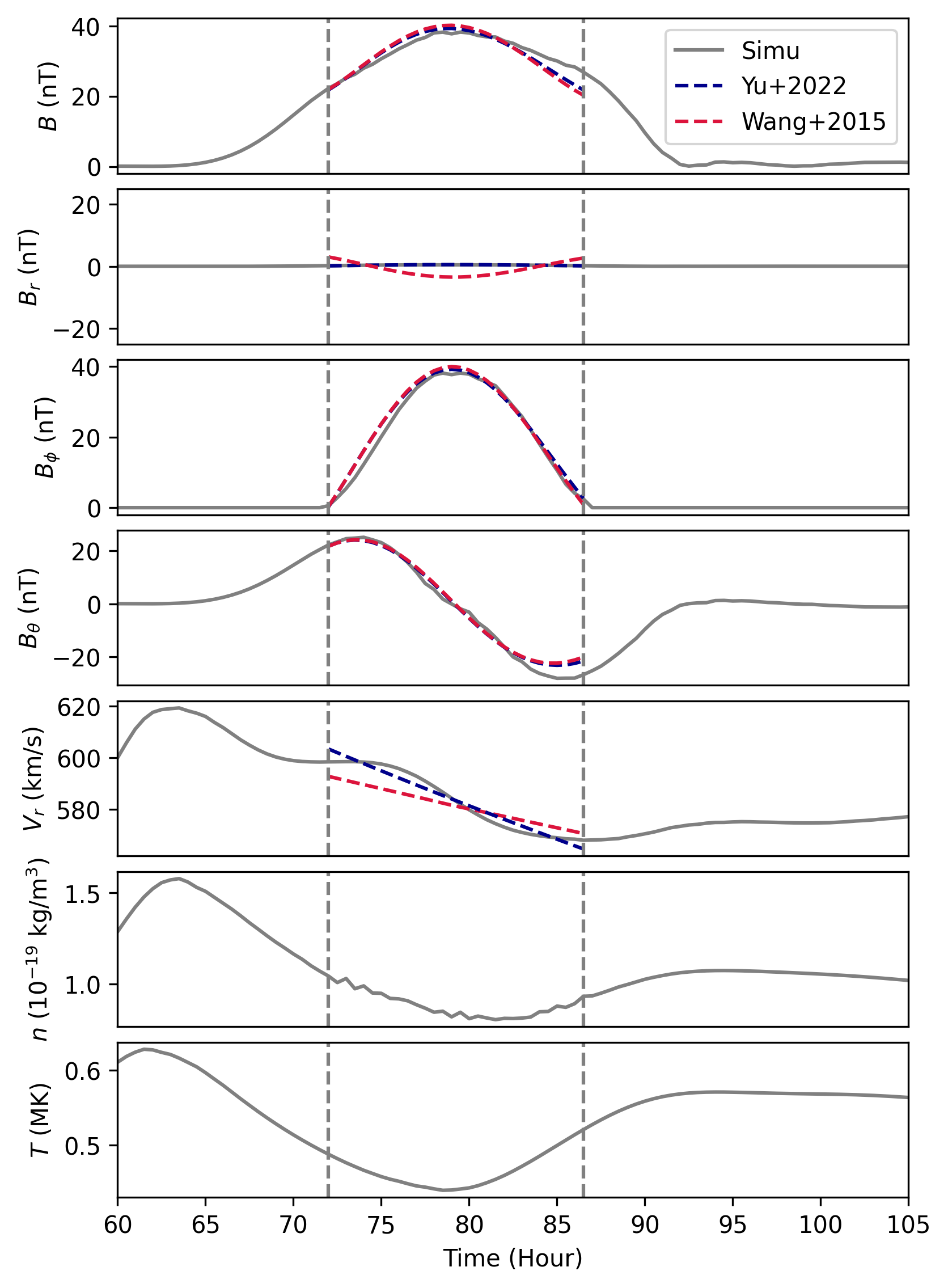}
    \caption{Synthetic in-situ temporal profiles. From top to bottom, the figure shows the magnetic field strength, three magnetic field components, radial proton velocity, proton density, and proton temperature. The vertical dashed lines mark the MFR boundaries. The red and blue dashed curves represent the fitted profiles based on the models of \cite{wang2015} and \cite{yu2022}, respectively. The blue dashed line in the velocity panel indicates the fit to the velocity-time data points using Equation~\ref{farrugia_eq}.
    }
    \label{fig: insitu}
\end{figure}
Figure~\ref{fig: insitu} shows the synthetic in-situ profiles of the magnetic field strength, three magnetic field components, radial proton velocity, proton density, and proton temperature from top to bottom based on the synthetic spacecraft at 215~$R_\odot$ (1 au) at $\Delta \theta=0^\circ$. The MFR boundaries are marked by the vertical dashed lines. Classic signatures of MCs are found inside the MFR region: enhanced magnetic field strength, rotation of the magnetic field vectors, and decrease in the proton temperature \citep{burlaga1981}. The axial magnetic field ($B_\phi$) outside the MFR region is zero. The decrease in the radial velocity from the front to the rear is one of the in-situ signatures indicating the existence of the radial expansion of the MFR \citep{farrugia1993}. Additionally, the duration of the MFR region is found to be around 18~hours, which is consistent with the average of 20~hours of magnetic clouds measured near 1~au \citep{richardson2010,regnault2020}. We note that, due to the lower grid resolution and smoothing treatment in simulations, the potential formation of the shock structure is not observed. The CME is expected to drive a shock when it is close to the Sun as the leading edge speed is faster than the fast magnetosonic speed in the solar wind frame. As the CME propagates, the CME leading edge decelerates, and a shock would not form or would not be driven anymore.

The two linear force-free MFR models are then applied to the synthetic in-situ parameters. The fitted profiles are shown by the red \citep{wang2015} and blue \citep{yu2022} dashed curves in Figure~\ref{fig: insitu}, and the fitted parameters are listed in Table~\ref{tb: table}. Note that the blue dashed line in the velocity panel indicates the fit to the velocity-time data points using Equation~\ref{farrugia_eq}. It is found that the fitted profiles match well the synthetic profile except for $B_r$ for the model of \cite{wang2015} (caused by a slight inconsistency of the fitted $\phi$). The results reveal the effectiveness of applying the linear force-free models to the simulated cases. The fitted profiles for the synthetic measurements at 15, 40, 80, 120, 160, and 260~$R_\odot$ at $\Delta \theta=0^\circ$ are shown in Figure~\ref{fig: app_otherfit1}, and the fitted profiles at $\Delta \theta=4^\circ$ are shown in Figure~\ref{fig: app_otherfit2}. The fitted profiles are found to generally match the synthetic profiles. The model of \cite{yu2022} fits well all three components, while the model of \cite{wang2015} fits well $B_\theta$ and $B_\phi$ but leads to slight inconsistencies of $B_r$.

\subsection{Comparison between Simulations and Fits} \label{sec: simu_vs_fit}
\begin{figure}[!hbt]
    \centering
    \includegraphics[width=1.\textwidth]{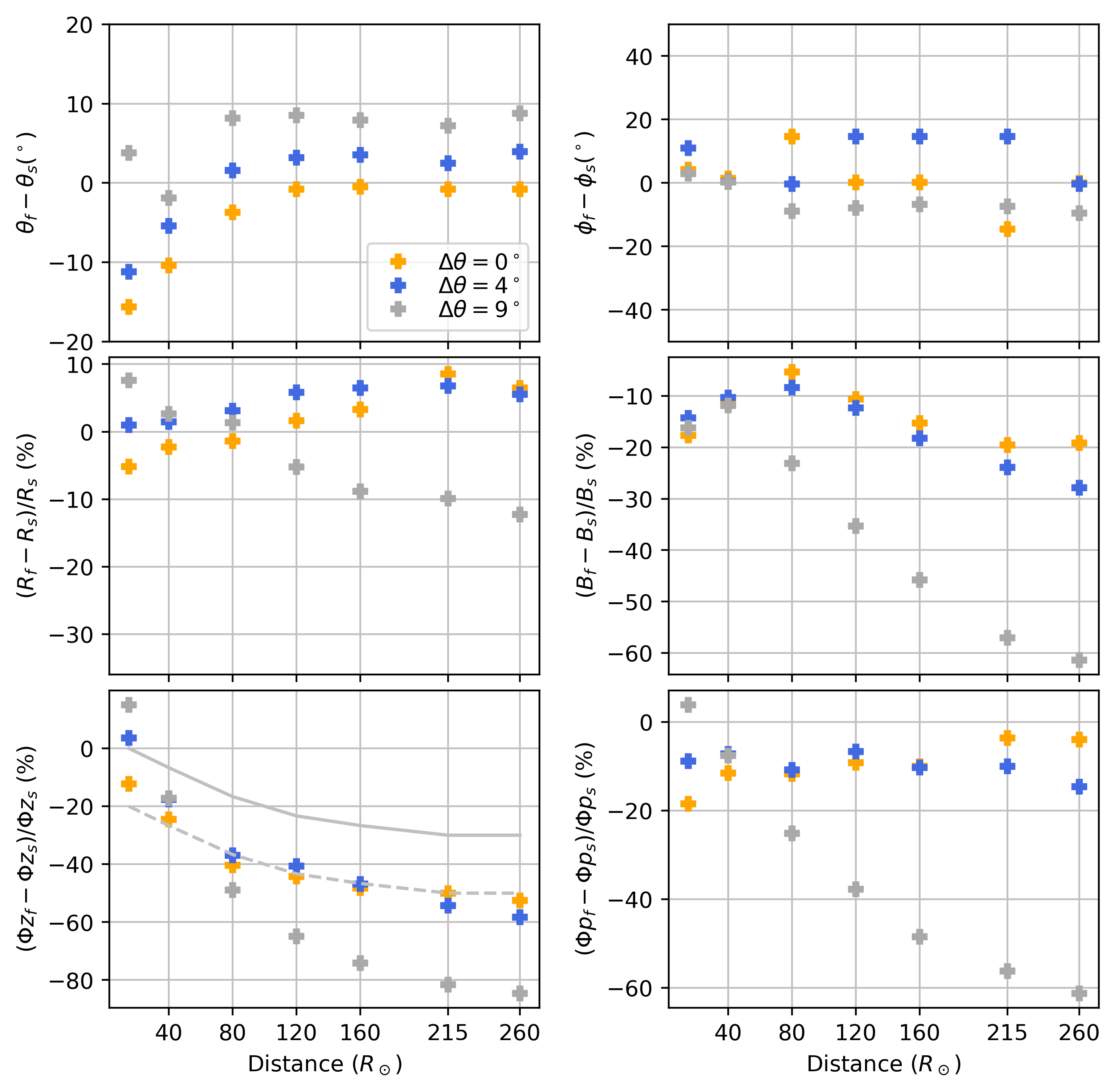}
    \caption{Comparisons between the simulation and fitted parameters using the model of \cite{wang2015}. The parameters include $\theta_0$, $\phi_0$, $R_0$, $B_0$, $\Phi_z$, and $\Phi_p$, and the subscripts ``f'' and ``s'' refer to the fitted and simulated parameters, respectively. The gray solid curve in the bottom left panel shows the relative difference between the axial magnetic flux in the central circular region and the total axial flux of the entire MFR. The gray dashed curve indicates a greater 20\% systematic underestimation compared to the solid curve.
    }
    \label{fig: fit_simu}
\end{figure}

We first investigate the fitting of the model of \cite{wang2015}. Figure~\ref{fig: fit_simu} shows the comparisons between the simulation and fitted parameters of the MFR, including $\theta_0$, $\phi_0$, $R_0$, $B_0$, $\Phi_z$, and $\Phi_p$. Subscripts ``f'' and ``s'' refer to the fitted and simulated parameters, respectively. Note that the simulated cases with the spacecraft trajectories at $\Delta \theta=4^\circ$ and $9^\circ$ are equivalent to the CME deflection by $4^\circ$ and $9^\circ$ relative to the equator, and thus $\theta_0$ under such conditions is $86^\circ$ and $81^\circ$, respectively.
We find that all the normalized $\chi$ are below 0.2, indicating that the fitted profiles are close to the synthetic in-situ profiles. There are four points to understand in the figure. First, except for the results at 15~$R_\odot$ and 40~$R_\odot$, $\theta_f$ is consistent with $\theta_s$ (differences within $\pm 10^\circ$). Second, the fitted $\phi$ is also consistent with that from the simulations, except for the two fits at $\Delta \theta=0^\circ$ and three at $\Delta \theta=4^\circ$ (but they are within $\pm 20^\circ$). Third, the fitted $R_0$ is roughly consistent with that in simulations with the differences being within $\pm \sim 10\%$. At $\Delta \theta=4^\circ$, the fitting overestimates $R_0$. At $\Delta \theta=0^\circ$, the relative difference varies monotonically from negative to positive, whereas the trend is reversed at $\Delta \theta=9^\circ$. Fourth, the fitting underestimates the magnetic field strength ($B_0$) at the MFR axis by around 10--20\% at $\Delta \theta=0^\circ$ and $4^\circ$, respectively. The underestimation becomes more significant at $\Delta \theta=9^\circ$ and shows a clear trend with distance.

We next turn our focus on $\Phi_z$ and $\Phi_p$, the axial and poloidal magnetic fluxes. Based on the bottom two panels, the fitting model underestimates both $\Phi_z$ and $\Phi_p$ for almost all crossings. First, we find that the underestimation of $\Phi_z$ becomes more significant when the angular separation from the Sun-MFR axis line (or impact parameter) is larger. This also applies to $\Phi_p$. We do not consider the results at $\Delta \theta=9^\circ$ in the following part but discuss its implication in Section~\ref{dis: unc}. We emphasize that a) the underestimation of $\Phi_z$ is distance-dependent, i.e., being greater at larger distances, and b) the underestimation of $\Phi_p$ is independent of distance. Using the results at $\Delta \theta=0^\circ$, the underestimation of $\Phi_z$ ranges from 12\% to 52\% when the MFR propagates from 15~$R_\odot$ to 215~$R_\odot$ and beyond. The results at $\Delta \theta=4^\circ$ show a very similar trend but the fit at 15~$R_\odot$ is consistent with the simulated $\Phi_z$. For both of these $\Delta \theta$, the underestimation does not change significantly beyond the MFR distance of 160~$R_\odot$. This is related to the fact that the MFR aspect ratio also remains almost constant beyond that distance (see the bottom right panel of Figure~\ref{fig: aspect_ratio}). As for $\Phi_p$, the underestimation is weaker compared to that of $\Phi_z$ (relative differences are within around $-20\%$), indicating that the fitted $\Phi_p$ is approximately consistent with the simulations. Furthermore, the fitted $\Phi_p$ is distance-independent (i.e., independent of the aspect ratio).

When the MFR is close to the Sun with its aspect ratio being close to 1, there exists an underestimation of $\Phi_z$ of $\sim$12\% at $\Delta \theta=0^\circ$, and the underestimation of $\Phi_p$ is also around $\sim$10--20\% at different distances. We illustrate here that such underestimations of $\Phi_z$ and $\Phi_p$ are systematic. Based on Equations~\ref{lundquist_eq4} and \ref{lundquist_eq5}, $\Phi_z$ and $\Phi_p$ in the Lundquist solution are determined by the fitted $B_0$ and $R_0$ ($\propto B_0 R_0^2$ for $\Phi_z$, and $\propto B_0 R_0$ for $\Phi_p$). Therefore, the underestimation of the fitted $B_0$ and the slight difference of $R_0$ as shown in the two middle panels of Figure~\ref{fig: fit_simu} can explain the systematic underestimations of $\Phi_z$ and $\Phi_p$. However, the underestimations of the fitted $B_0$ and $R_0$ are distance-independent (at $\Delta \theta=0^\circ$ and $4^\circ$), and thus the distance-dependent additional underestimation of $\Phi_z$ requires further elucidation, which is attributed to the model's circular cross-section assumption and is discussed in Section~\ref{sec: dis}.

\begin{figure}[!hbt]
    \centering
    \includegraphics[width=1.\textwidth]{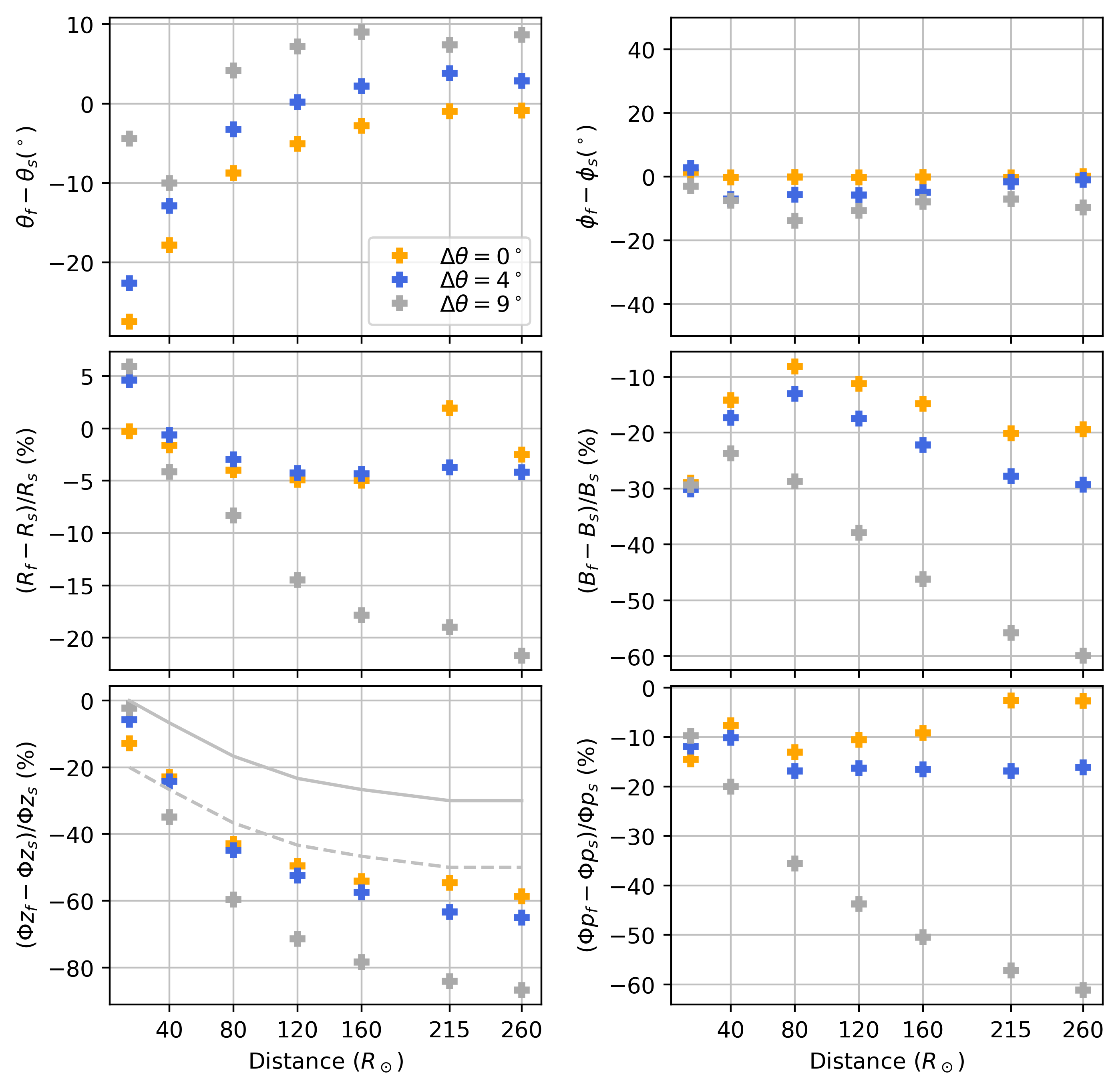}
    \caption{Similar to Figure~\ref{fig: fit_simu} but using the model of \cite{yu2022}.
    }
    \label{fig: fit_simu_yu}
\end{figure}
We now turn our attention to the fitted results of the model of \cite{yu2022}. Figure~\ref{fig: fit_simu_yu} shows the comparisons of the parameters obtained from simulations and fittings. Due to the incorporation of the MVA method, it is found that this model can lead to consistent $\phi_0$ at $\Delta \theta=0^\circ$ (at $\Delta \theta=4^\circ$, the difference is still small). In general, the comparisons as shown in Figure~\ref{fig: fit_simu_yu} are similar to those for the model of \cite{wang2015} in Figure~\ref{fig: fit_simu}. First, most of the fitted $R_0$ and $B_0$ are slightly smaller. Second, the fitted $\Phi_z$ at $\Delta \theta=0^\circ$ and $4^\circ$ are underestimated compared to $\Phi_z$ in simulations, and this underestimation is found to be distance-dependent. Third, the fitted $\Phi_p$ at $\Delta \theta=0^\circ$ and $4^\circ$ are slightly smaller (distance-independent) than the simulated $\Phi_p$. Last, the fitted $R_0$, $B_0$, $\Phi_z$, and $\Phi_p$ at $\Delta \theta=9^\circ$ are significantly inconsistent with the MFR parameters in simulations.

\section{Discussion} \label{sec: dis}
In this section, we focus on the reason for the underestimation of the axial magnetic flux using a circular cross-section model to fit an elliptical MFR and discuss the uncertainties and implications related to the model's cross-section and the spacecraft's impact parameter.

\subsection{Numerical Explanation} \label{dis: explain}
\begin{figure}[!hbt]
    \centering
    \includegraphics[width=1\textwidth]{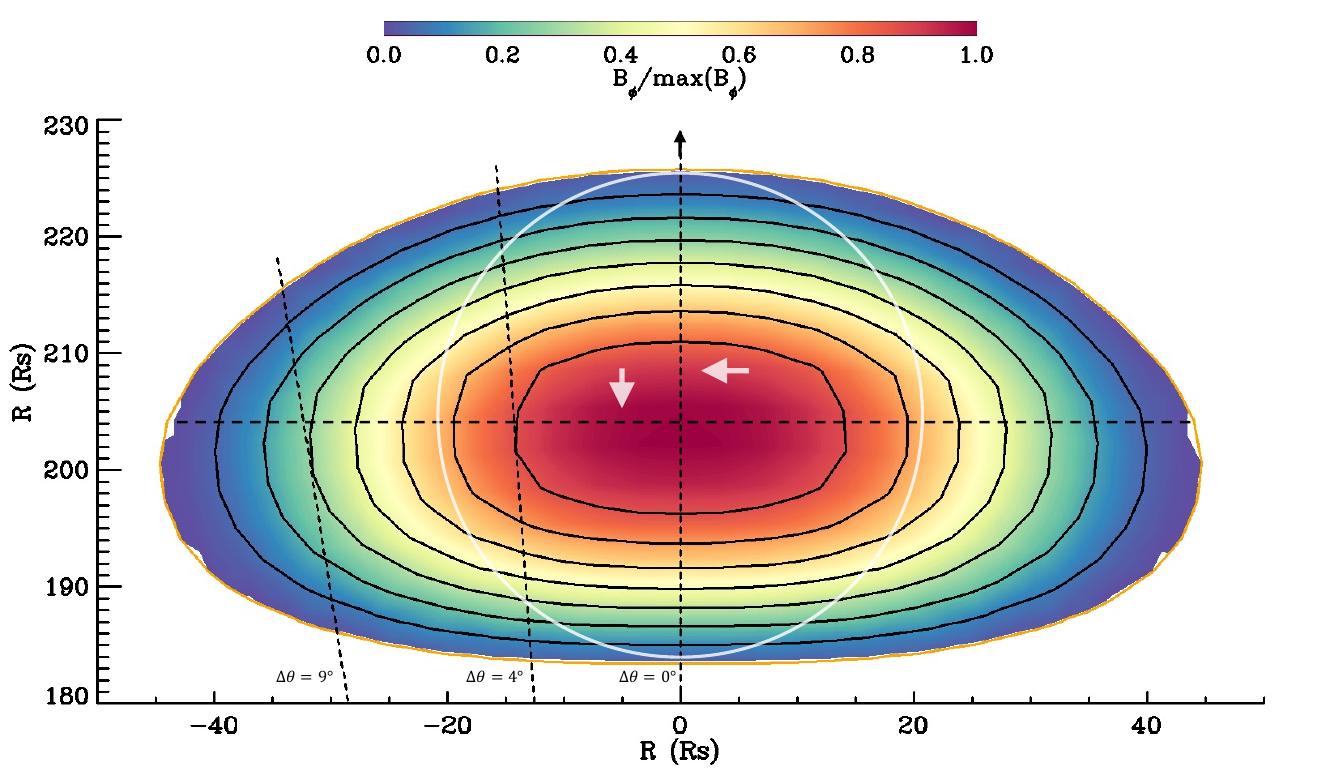}
    \caption{Schematic of using a circular cross-section model to fit an elliptical MFR. The MFR is obtained from the simulations when the MFR front propagates to around 215~$R_\odot$. The color map indicates the ratio of the axial magnetic field ($B_\phi$) strength to the maximum axial field strength. The black arrow indicates the MFR propagation direction, and the three synthetic spacecraft trajectories at $\Delta \theta=0^\circ$, $4^\circ$, and $9^\circ$ are shown by the black dashed lines. The two white arrows indicate the poloidal magnetic flux integrated along the horizontal and vertical dashed lines. This MFR is taken at the same time as that in the bottom panel of Figure~\ref{fig: mfrcontour}.
    }
    \label{fig: numerical_schematic}
\end{figure}

We use Figure~\ref{fig: numerical_schematic} to illustrate the underestimation of $\Phi_z$ as a function of distance. In this figure, the MFR is obtained from our simulations when the MFR front reaches about 215~$R_\odot$ (1~au). The aspect ratio of the MFR at that distance is around two. The color map indicates the ratio of the axial magnetic field strength to the maximum axial field strength. It is found that the majority of the axial magnetic field is located in the region near the MFR center (axis), which is also supported by the synthetic in-situ profiles in the radial direction. The black arrow and the vertical dashed line indicate the MFR propagation direction, and the three synthetic spacecraft trajectories at $\Delta \theta=0^\circ$, $4^\circ$, and $9^\circ$ are shown by the black dashed lines. If we use a model with a circular cross-section to fit the in-situ data, then the model only covers the central region of the MFR and misses the outer part, indicated by the white circle in the figure (note that the trajectory at $\Delta \theta=9^\circ$ is outside the central circular region). Therefore, even though the axial magnetic field can be well-fitted, the total axial magnetic flux is still underestimated because the axial field (flux) outside the central circle is not considered. The more flattened the MFR is, the larger the underestimation may be. This is further confirmed in simulations. 

We calculate the axial magnetic flux just within a central circular region of the MFR (i.e., the white circle), with a diameter spanning from the front to the rear (i.e., the vertical dashed line), at the six selected different distances. The fraction is calculated by $(\Phi_{zc}-\Phi_z)/\Phi_z$, where $\Phi_{zc}$ is the axial flux within the circular region and $\Phi_z$ is the total flux. The solid curve in the bottom left panel of Figures~\ref{fig: fit_simu} and \ref{fig: fit_simu_yu} displays the variation of the fraction, showing that the fraction decreases with distance and becomes approximately constant of around 20\% after around 150~$R_\odot$. Note that since most of the strong $B_z$ is distributed in the region near the axis, the underestimation at a larger aspect ratio is not very significant. If we add an upper systematic underestimation of around 20\%, based on the underestimations of $B_0$ and $R_0$, while noting that $\Phi_z$ is proportional to $B_0R_0^2$ (see Section~\ref{sec: simu_vs_fit}), to the solid curve to derive the dashed curve in Figures~\ref{fig: fit_simu} and \ref{fig: fit_simu_yu}, we find that the dashed curve is consistent with the estimates at $\Delta \theta =0^\circ$ and 4$^\circ$ as shown in the figures. Therefore, it further supports the conclusion that the additional underestimation of $\Phi_z$ is caused by the deformation of the MFR.

We also discuss the estimation of the poloidal magnetic flux $\Phi_p$. In numerical simulations, $\Phi_p$ of the MFR is calculated using the difference between the magnetic flux function values obtained at the MFR axis (where it has the maximum value) and at the MFR boundary. Such a difference is independent of the deformation of the MFR. Moreover, in the fitting models, $\Phi_p$ can be estimated by integrating the poloidal magnetic field across half of the vertical (at $\Delta \theta=0^\circ$) or horizontal dashed line (also indicated by the white arrows in Figure~\ref{fig: numerical_schematic}) inside the MFR. Integral along either the vertical or horizontal line is conserved and independent of the shape of the MFR. Therefore, the estimation of $\Phi_p$ is independent of the distance (or aspect ratio).

\subsection{Uncertainties and Implications} \label{dis: unc}
In our simulations, the mass and magnetic fluxes are conserved during the MFR propagation. However, in reality, CMEs may experience internal magnetic reconnection \citep{myers2015} or magnetic erosion \cite[e.g.,][]{dasso2006,ruffenach2012,lavraud2014,wang2018,farrugia2023} during their eruption and propagation. Internal reconnection can convert axial magnetic flux to poloidal magnetic flux but the MFR helicity remains conserved if considering the MFR as a closed system. Magnetic erosion can peel off the outer shells of the MFR and thus cause decreases in the magnetic fluxes and mass of the MFR. These two aspects are out of the scope of this study.

We discuss the influence of the asymmetry of the in-situ magnetic field profiles on the fitted parameters. As shown in Figures~\ref{fig: app_otherfit1} and \ref{fig: app_otherfit2}, the MFR magnetic profiles are more asymmetric (stronger at the front) close to the Sun, while, with increasing solar distances, these profiles become more symmetric. It is well known that the aging effect \citep{osherovich1993,demoulin2020,regnault2024b} in association with the CME expansion can result in such asymmetric magnetic profiles. However, Figures~\ref{fig: app_otherfit1} and \ref{fig: app_otherfit2} with well fitted expansion speeds indicate that the incorporation of the MFR expansion is not sufficient to explain the asymmetric magnetic profiles when the MFR is close to the Sun \citep[similar to the results as shown in][]{yu2022}. The compression from the stretched overlying coronal magnetic field ahead of the MFR may provide additional contributions to the stronger field magnitude at the front. Such an asymmetry leads to uncertainties in the fitted MFR axis orientation, especially the elevation angle $\theta_f$, which may further affect other fitted parameters. Based on Equations~\ref{lundquist_eq4} and \ref{lundquist_eq5}, conservation of the magnetic fluxes depends on the fitted $B_0$ and $R_0$. Within the limitation of the two circular cross-section models, we test the fitted $B_0$ and $R_0$ for the case at 15~$R_\odot$ and $\Delta \theta=0^\circ$ by fixing $\theta_f=90^\circ$ in the models. We find that, in the model of \citet{wang2015}, $B_0$ and $R_0$ are not significantly affected but the fitted magnetic field profiles become more inconsistent with the simulations; in the model of \citet{yu2022}, $B_0$ is underestimated while $R_0$ remains approximately the same (indicating that the impact parameter is not affected). We note that those results are obtained based on a fitting to minimize  $\chi$; however, the minimization fitting technique may result in the MFR parameters that are inconsistent with the true parameters observed in simulations, which need further investigation in the future. In addition, the aging effect associated with the MFR acceleration during its propagation can also influence the asymmetry of the in-situ magnetic field profiles.

We then discuss the conservation of the MFR magnetic helicity here, as helicity is one of the fundamental parameters of MFRs. In our numerical method, even though it solves 2.5-D problems, we can also calculate the magnetic helicity per unit length, which depends on the axial magnetic field and magnetic flux function \cite[][and see their Equation~33]{hu1997}. Such expression is theoretically conserved and has been numerically confirmed to be conserved in the presence of reconnection \citep{hu1997}. In our simulations, the total magnetic helicity of the MFR is also conserved. In the Lundquist solution model, magnetic helicity is expressed as follows:
\begin{equation} \label{lundquist_eq6}
    H_m = \int_{0}^{l} \int_{0}^{2\pi} \int_{0}^{R_0} \bm{A} \cdot \bm{B} r dr d\varphi dz = 0.701 B_0^2R_0^3l,
\end{equation}
in which $\bm{A}$ is a vector potential of $\bm{B}$. Equation~\ref{lundquist_eq6} reveals that $H_m$ is proportional to $\Phi_z \times \Phi_p$. Thus, the MFR helicity will also be underestimated, and such underestimation depends on the MFR aspect ratio.

Overall, we have shown that, although the linear force-free circular cross-section model can fit the synthetic in-situ profiles well, it can lead to an underestimation of the axial magnetic flux of the flattened MFR but obtain relatively consistent poloidal flux. The cases with nonlinear force-free or even non-force-free magnetic field configurations need to be further investigated. In addition, in Figures~\ref{fig: fit_simu} and \ref{fig: fit_simu_yu}, the more significant underestimations of $\Phi_z$ and $\Phi_p$ at larger $\Delta \theta$ are due to the underestimation of $B_0$. Combining Figure~\ref{fig: numerical_schematic}, we deduce that the detection of the MFR structure is still available for a flattened MFR even though the spacecraft trajectory is distant from the MFR axis (with larger impact parameters); however, since the majority of the strong axial magnetic field is distributed in the inner circular region of the MFR for a deformed MFR, the fitting using a circular cross-section model may not well obtain the true $B_0$. The incorporation of the models with non-circular (elliptical) cross-sections may be helpful to resolve this problem.
Since there are varieties of non-circular cross-section models with different axis symmetry assumptions, we plan to conduct these studies with different models in the future.

\section{Conclusion} \label{sec: con}
Magnetic fluxes (poloidal and axial) of MFR can be used as good indicators to connect the properties of CMEs estimated in the corona and in-situ in interplanetary space, as has been done in past work \citep[e.g.,][]{qiu2007,mostl2009,hu2014}. In the ideal MHD frame, magnetic fluxes are conserved. We use 2.5-D MHD numerical simulations to study the influence of the MFR deformation on the magnetic fluxes of MFR estimated in-situ using a circular cross-section MFR model. In simulations, the poloidal and axial magnetic fluxes of the MFR are conserved during its propagation. The cross-section of the MFR becomes flattened, and its aspect ratio (lateral size to radial size) increases to around two when the MFR propagates beyond the distance of around 150~$R_\odot$. Analyses of in-situ measurements have revealed that the MFR aspect ratio may be of the order of two to three near one au, consistent with the simulation used here \citep[]{demoulin2013,lugaz2024}. 

We create synthetic spacecraft measurements at different distances, i.e., at 15, 40, 80, 120, 160, 215, and 260~$R_\odot$, which cover simulated MFR conditions with different aspect ratios. We use two cylindrical linear force-free MFR models with a circular cross-section \citep{wang2015,yu2022} to fit the synthetic in-situ data. We find that the fitting underestimates the axial magnetic flux, and such underestimation depends on the distance (or aspect ratio); the more flattened the MFR becomes, the more significant the underestimation is. This is because a circular cross-section only considers the axial magnetic field in the central circular region and thus neglects the axial field outside the central region. In our simulations, the central circular region is found to contain $\sim$80\% of the total axial magnetic flux when the aspect ratio of the MFR cross-section is around two. In fittings with circular cross-section models, an elliptical cross-section with an aspect ratio of two leads to a 50\% underestimation in the axial flux (a systematic underestimation in the fitted axis magnetic field strength is incorporated). Based on the events measured near 1~au, \cite{demoulin2013} found that the aspect ratio of the MFR is elliptical with an aspect ratio of two or three. However, the fitting of the poloidal flux is found to be independent of the deformation of the MFR when the synthetic spacecraft impact parameter is small or moderate. The poloidal flux is related to the integral of the poloidal magnetic field in the plane of the MFR cross-section along a line from the axis to the boundary of the MFR. Such integral is conserved regardless of how the MFR cross-section changes. These results further indicate that we need to take care of the uncertainties when using circular cross-section MFR models to get the MFR properties in heliophysics and space weather studies. Furthermore, for the next step, we aim to conduct studies using elliptical cross-section models.

\section*{Acknowledgments}
\noindent{\bf{Author Contributions}} B.Z. led the study and performed the analysis, and wrote the manuscript. All authors discussed and contributed to the analysis and revised the manuscript.

\noindent{\bf{Funding}} B.~Z. acknowledges the NASA ECIP grant 80NSSC23K1057. B.~Z. and N.~L. acknowledge the NASA grants 80NSSC17K0009 and 80NSSC20K0700, and the NSF grant AGS-2301382. N.A. acknowledges support from NSF AGS-1954983 and NASA grant 80NSSC22K0349 and NASA ECIP 80NSSC21K0463. R.M.~W acknowledges the NASA grant 80NSSC19K0914. E.E.~D., U.V.~A., and H.T.~R. are funded by the European Union (ERC, HELIO4CAST, 101042188). Views and opinions expressed are however those of the author(s) only and do not necessarily reflect those of the European Union or the European Research Council Executive Agency. Neither the European Union nor the granting authority can be held responsible for them. 

\noindent{\bf{Data Availability}} The simulated data and fitted results can be found on Zenodo (\url{https://doi.org/10.5281/zenodo.14884689}).

\section*{Declarations}
\noindent{\bf{Competing interests}} The authors declare no competing interests.

\begin{appendix}
\section{Appendix}
\setcounter{figure}{0}
\renewcommand{\thefigure}{A\arabic{figure}}
Figures~\ref{fig: app_otherfit1} and \ref{fig: app_otherfit2} show the simulated magnetic field vectors and radial velocity at different distances and at $\Delta \theta=0^\circ$ and $4^\circ$, overlaid by the fitted results of the models of \cite{wang2015} and \cite{yu2022}, respectively. The blue dashed lines in the velocity panels indicate the linear fit to the velocity-time measurements. It is found that the fitted profiles are consistent with the simulation profiles except for the minor inconsistencies of $B_r$ using the model of \cite{wang2015}.
\begin{figure}[!hbt]
    \centering
    \includegraphics[width=1.\textwidth]{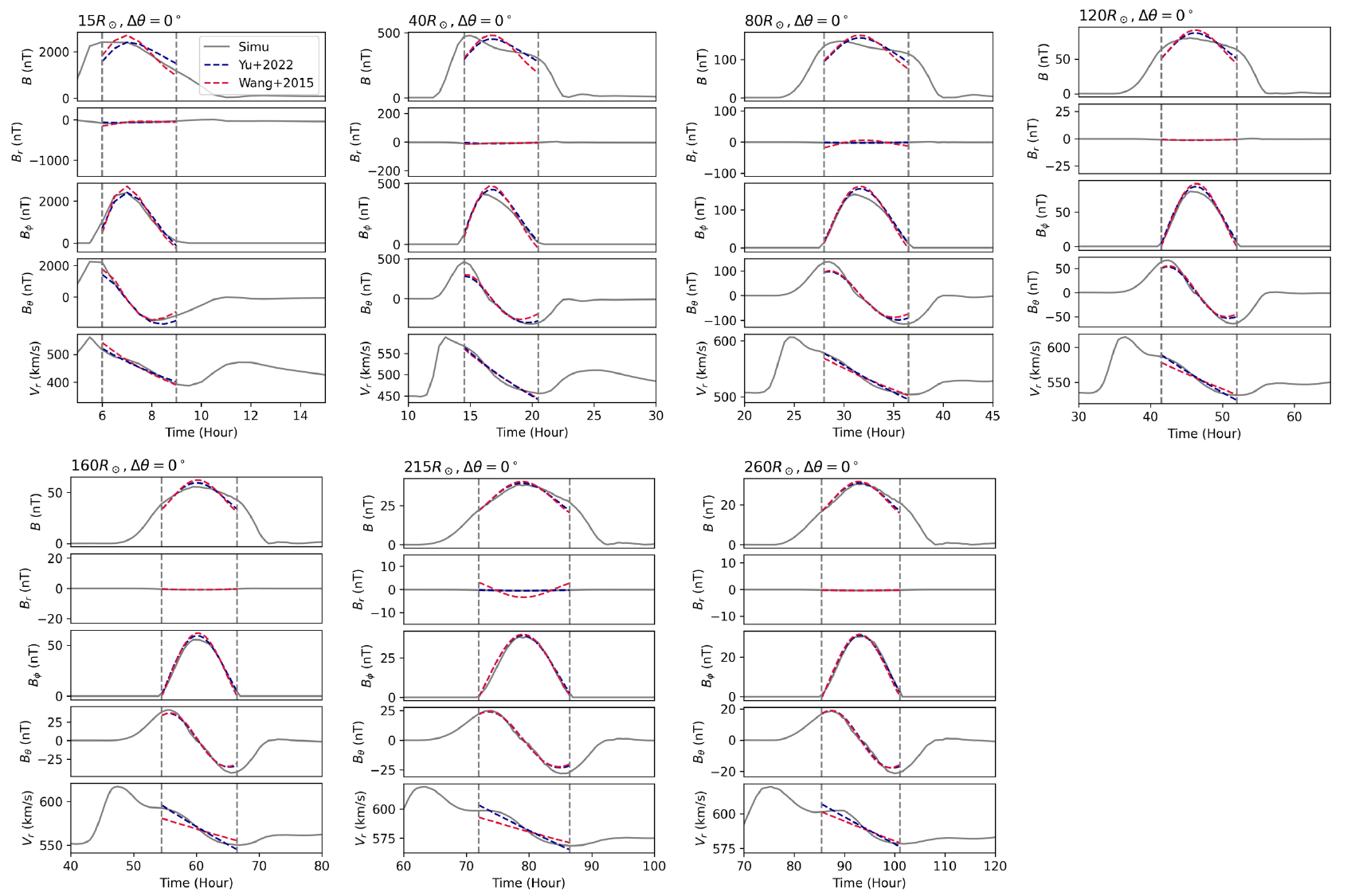}
    \caption{Simulated magnetic field vectors and radial velocity (gray solid curve) at different distances and at $\Delta \theta=0^\circ$, overlaid by the fitted results (dashed curves).
    }
    \label{fig: app_otherfit1}
\end{figure}
\begin{figure}[!hbt]
    \centering
    \includegraphics[width=1.\textwidth]{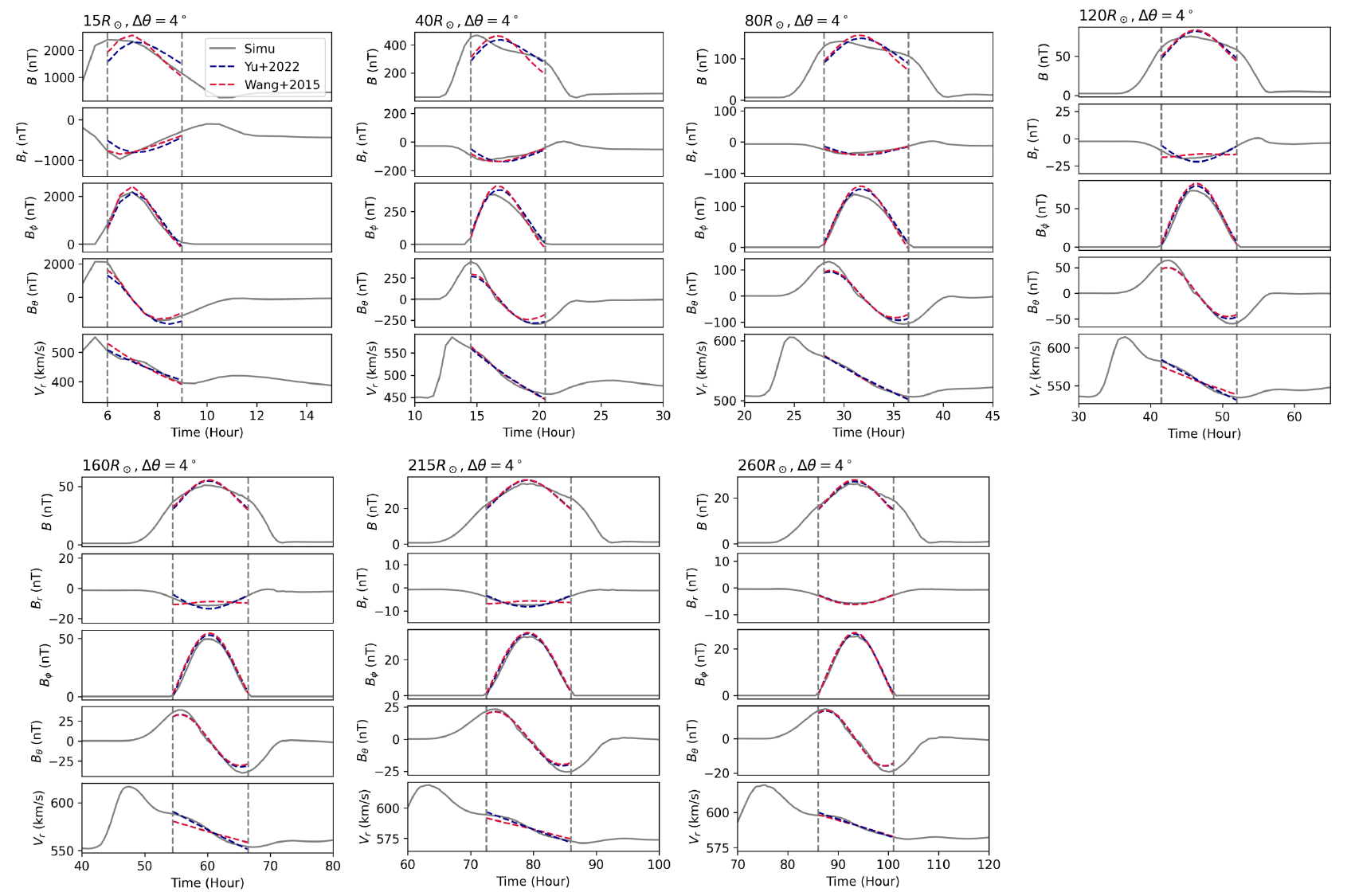}
    \caption{Simulated magnetic field vectors and radial velocity (gray solid curve) at different distances and at $\Delta \theta=4^\circ$, overlaid by the fitted results (dashed curves).
    }
    \label{fig: app_otherfit2}
\end{figure}
\end{appendix}

\clearpage

\begin{sidewaystable}
\centering
\caption{CME parameters from simulations and fittings of the models of \cite{wang2015} and \cite{yu2022}}
\footnotesize
\begin{tabular}{|c|c|c|c|c|c|c|c|c|c|c|c|c|c|c|c|c|c|c|c|}
\hline
SC & SC & \multicolumn{6}{c|}{Simulations} & \multicolumn{6}{c|}{Model of \cite{wang2015}} & \multicolumn{6}{c|}{Model of \cite{yu2022}} \\ \cline{3-20}
Dis & $\Delta \theta$ & $R_0$ & $B_0$ & ($\theta_0, \phi_0$) & $d$ & $\Phi_z$ & $\Phi_p$ & $R_0$ & $B_0$ & ($\theta_0, \phi_0$) & $d$  & $\Phi_z$ & $\Phi_p$ & $R_0$ & $B_0$ & ($\theta_0, \phi_0$) & $d$  & $\Phi_z$ & $\Phi_p$ \\ \hline
~ & $0^\circ$ & & & ($90^\circ$, $90^\circ$) & 0.00 & & & 3.0 & 3560 & ($75^\circ$, $94^\circ$) & 0.18 & 2.13 & 0.85 & 3.2 & 3076 & ($63^\circ$, $91^\circ$) & 0.04 & 2.04 & 0.89 \\ 
15 & $4^\circ$ & 3.2 & 4328 & ($86^\circ$, $90^\circ$) & 0.30 & 2.42 & 1.04 & 3.2 & 3710 & ($75^\circ$, $101^\circ$) & 0.43 & 2.51 & 0.94 & 3.3 & 3022 & ($64^\circ$, $93^\circ$) & 0.30 & 2.21 & 0.92 \\ 
~ & $9^\circ$ & & & ($81^\circ$, $90^\circ$) & 0.67 & & & 3.4 & 3626 & ($85^\circ$, $93^\circ$) & 0.67 & 2.78 & 1.08 & 3.4 & 3057 & ($77^\circ$, $87^\circ$) & 0.64 & 2.29 & 0.94 \\ \hline
%
~ & $0^\circ$ & & & ($90^\circ$, $90^\circ$) & 0.00 & & & 6.9 & 589 & ($80^\circ$, $91^\circ$) & 0.06 & 1.83 & 0.92 & 7.0 & 572 & ($72^\circ$, $90^\circ$) & 0.02 & 1.81 & 0.96 \\ 
40 & $4^\circ$ & 7.1 & 665 & ($86^\circ$, $90^\circ$) & 0.28 & 2.42 & 1.04 & 7.2 & 597 & ($81^\circ$, $90^\circ$) & 0.25 & 2.00 & 0.97 & 7.0 & 550 & ($73^\circ$, $83^\circ$) & 0.16 & 1.78 & 0.93 \\ 
~ & $9^\circ$ & & & ($81^\circ$, $90^\circ$) & 0.64 & & & 7.3 & 586 & ($79^\circ$, $90^\circ$) & 0.59 & 2.01 & 0.96 & 6.8 & 508 & ($71^\circ$, $82^\circ$) & 0.53 & 1.52 & 0.83 \\ \hline
%
~ & $0^\circ$ & & & ($90^\circ$, $90^\circ$) & 0.00 & & & 10.9 & 188 & ($86^\circ$, $105^\circ$) & 0.18 & 1.45 & 0.92 & 10.6 & 183 & ($81^\circ$, $90^\circ$) & 0.01 & 1.34 & 0.90 \\ 
80 & $4^\circ$ & 11.0 & 199 & ($86^\circ$, $90^\circ$) & 0.29 & 2.42 & 1.04 & 11.4 & 183 & ($88^\circ$, $90^\circ$) & 0.21 & 1.53 & 0.93 & 10.7 & 173 & ($83^\circ$, $84^\circ$) & 0.15 & 1.30 & 0.86 \\ 
~ & $9^\circ$ & & & ($81^\circ$, $90^\circ$) & 0.66 & & & 11.2 & 153 & ($89^\circ$, $81^\circ$) & 0.47 & 1.24 & 0.78 & 10.1 & 142 & ($85^\circ$, $76^\circ$) & 0.43 & 0.95 & 0.67 \\ \hline
%
~ & $0^\circ$ & & & ($90^\circ$, $90^\circ$) & 0.00 & & & 14.5 & 99 & ($90^\circ$, $90^\circ$) & 0.02 & 1.35 & 0.95 & 13.6 & 98 & ($85^\circ$, $90^\circ$) & 0.01 & 1.18 & 0.93 \\ 
120 & $4^\circ$ & 14.3 & 76 & ($86^\circ$, $90^\circ$) & 0.30 & 2.42 & 1.04 & 15.1 & 97 & ($89^\circ$, $105^\circ$) & 0.34 & 1.44 & 0.97 & 13.6 & 91 & ($86^\circ$, $84^\circ$) & 0.13 & 1.11 & 0.87 \\ 
~ & $9^\circ$ & & & ($81^\circ$, $90^\circ$) & 0.67 & & & 13.5 & 72 & ($89^\circ$, $82^\circ$) & 0.48 & 0.86 & 0.65 & 12.2 & 69 & ($88^\circ$, $79^\circ$) & 0.46 & 0.69 & 0.59 \\ \hline
%
~ & $0^\circ$ & & & ($90^\circ$, $90^\circ$) & 0.00 & & & 17.3 & 65 & ($90^\circ$, $90^\circ$) & 0.02 & 1.26 & 0.94 & 15.9 & 65 & ($87^\circ$, $90^\circ$) & 0.01 & 1.08 & 0.94 \\ 
160 & $4^\circ$ & 16.8 & 111 & ($86^\circ$, $90^\circ$) & 0.29 & 2.42 & 1.04 & 17.8 & 62 & ($90^\circ$, $105^\circ$) & 0.33 & 1.29 & 0.94 & 16.0 & 59 & ($88^\circ$, $85^\circ$) & 0.13 & 1.00 & 0.87 \\ 
~ & $9^\circ$ & & & ($81^\circ$, $90^\circ$) & 0.66 & & & 15.3 & 41 & ($89^\circ$, $83^\circ$) & 0.52 & 0.63 & 0.54 & 13.8 & 42 & ($90^\circ$, $82^\circ$) & 0.51 & 0.51 & 0.52 \\ \hline
%
~ & $0^\circ$ & & & ($90^\circ$, $90^\circ$) & 0.00 & & & 21.0 & 43 & ($90^\circ$, $75^\circ$) & 0.14 & 1.21 & 1.00 & 19.7 & 42 & ($89^\circ$, $90^\circ$) & 0.01 & 1.06 & 1.01 \\ 
215 & $4^\circ$ & 19.3 & 53 & ($86^\circ$, $90^\circ$) & 0.28 & 2.42 & 1.04 & 20.6 & 40 & ($89^\circ$, $105^\circ$) & 0.33 & 1.11 & 0.94 & 18.6 & 38 & ($90^\circ$, $88^\circ$) & 0.17 & 0.86 & 0.87 \\ 
~ & $9^\circ$ & & & ($81^\circ$, $90^\circ$) & 0.63 & & & 17.4 & 23 & ($88^\circ$, $83^\circ$) & 0.57 & 0.45 & 0.45 & 15.7 & 23 & ($88^\circ$, $83^\circ$) & 0.57 & 0.37 & 0.37 \\ \hline
%
~ & $0^\circ$ & & & ($90^\circ$, $90^\circ$) & 0.00 & & & 23.3 & 33 & ($90^\circ$, $90^\circ$) & 0.02 & 1.15 & 1.00 & 21.3 & 33 & ($90^\circ$, $90^\circ$) & 0.01 & 0.98 & 1.01 \\ 
260 & $4^\circ$ & 21.9 & 40 & ($86^\circ$, $90^\circ$) & 0.27 & 2.42 & 1.04 & 23.0 & 29 & ($90^\circ$, $90^\circ$) & 0.19 & 1.01 & 0.88 & 21.0 & 28 & ($89^\circ$, $89^\circ$) & 0.18 & 0.82 & 0.87 \\ 
~ & $9^\circ$ & & & ($81^\circ$, $90^\circ$) & 0.61 & & & 19.2 & 16 & ($90^\circ$, $80^\circ$) & 0.62 & 0.37 & 0.40 & 17.1 & 16 & ($90^\circ$, $80\circ$) & 0.62 & 0.31 & 0.31 \\ \hline
\end{tabular}
\begin{tablenotes}
\item[1] [1] The unit of the spacecraft distance is $R_\odot$, of $R_0$ is $R_\odot$, of $B_0$ is nT, of $\Phi_z$ is $\times10^{21}$~Mx, and of $\Phi_p$ is $\times10^{22}$~Mx.
\item[2] [2] Note that $d$ in simulations is the ratio of the closest distance between the spacecraft and the MFR axis to the half of the MFR lateral length. We do not consider the sign of $d$ due to the symmetry of the MFR with respective to the equator.
\item[3] [3] Although the MFR propagation is along $\theta=90^\circ$, the trajectories of the synthetic spacecraft along $\Delta \theta=4^\circ$ and $9^\circ$ are equivalent to the CME deflection by $4^\circ$ and $9^\circ$ while the spacecraft is still at $\theta=90^\circ$, and thus $\theta_0$ in simulations is $86^\circ$ and $81^\circ$, respectively.
\end{tablenotes}
\normalsize
\label{tb: table}
\end{sidewaystable}
\clearpage

\bibliographystyle{spr-mp-sola}

\end{document}